# $H_0$ and Odds on Cosmology

Andrew Jaffe

Canadian Institute for Theoertical Astrophysics,

60 St. George St., Toronto, Ontario M5S 1A1, Canada

## ABSTRACT

Recent observations by the Hubble Space Telescope of Cepheids in the Virgo cluster imply a Hubble Constant $H_0 = 80 \pm 17$ km/sec/Mpc. We attempt to clarify some issues of interpretation of these results for determining the global cosmological parameters $\Omega$ and $\Lambda$. Using the formalism of Bayesian model comparison, the data suggest a universe with a nonzero cosmological constant $\Lambda > 0$, but vanishing curvature: $\Omega + \Lambda = 1$.

*Subject headings:* Cosmology: theory – observation

What have we really learned from the recent observations of Virgo Cepheids and their implications for the Hubble Constant? That is, what have we learned about global cosmology from these observations, $H_0 = 80 \pm 17$ km/sec/Mpc from HST (Freedman et al. 1994), or $H_0 = 87 \pm 7$ km/sec/Mpc from CFHT (Pierce et al. 1994) combined with a lower limit on the age of the universe from stellar evolution of $t_0 \gtrsim 12$ Gyr? We shall use Bayesian statistics to hopefully give a precise answer to this question.

Bayes' theorem states
$$p(\theta|DI) = \frac{p(\theta|I)p(D|\theta I)}{p(D|I)} \tag{1}$$
where $p(a|bc)$ roughly means "the probability [density] of $a$ given $b$ and $c$. Here, $\theta$ represents the parameters of the theory we are considering (or more precisely, the statement that the parameters lie in some range), $D$ the outcome of some experiment, and $I$ any background information. Then $p(\theta|I)$ is the prior distribution for the parameters, $p(D|\theta I)$ is the likelihood of the data, and $p(D|I)$ is known as the *evidence*. Usually, this theorem is used to decide how the experiment effects our knowledge of the parameters of the theory. It can also be used in a more general context to compare theories and see which better explain the data (MacKay 1992).

This Bayesian approach to theory-testing has the advantage that it automatically incorporates Ockham's razor, favoring the simpler theory unless the more complicated one is significantly better at explaining the data. We let $j, k, \ldots$ represent the different models and write the background information corresponding to each as $I = I_j + I_k + \cdots$, where $+$ is "logical or". Then the likelihood of model $j$ is $p(D|jI) = p(D|I_j)$, just the evidence for the model as defined above. In this formalism, the ratio of the probabilities of two models (i.e., the "odds" favoring one or the other) is given by

$$\frac{p(j|DI)}{p(k|DI)} = \frac{p(j|I)}{p(k|I)} \frac{\int d\theta_j \, p(\theta_j|I_j)p(D|\theta_j I_j)}{\int d\theta_k \, p(\theta_k|I_k)p(D|\theta_k I_k)} \quad (2)$$

where $\theta_j$ and $I_j$ refer to the parameters and background information required for model $j$, and $p(j|I)$ is the prior probability of the model. We shall concentrate on the second factor (known as the *Bayes factor* which contains the experimental information. A model is favored by this factor if the average of its likelihood with respect to the prior distribution is greater—if more of its parameter space is likely, given the data. Thus, if there are large areas of the allowed parameter space with very low likelihoods, the model as a whole may be disfavored, even if it contains a strongly favored maximum likelihood.

The data that we consider are recent observations of the cepheids in the Virgo cluster, giving a value for the Hubble constant, and combinations of data and theory that give a lower limit on the age of the universe. We will consider the Hubble constant data to be represented by the *likelihood*

$$p(D|H_0 I) = N(H_0, \bar{H}, \delta H_0) \quad (3)$$

and the information about the age of the universe by the prior (we could equally well write this as a likelihood for some other set of data)

$$p(t_0|I) = L(t_0, \bar{t}, \delta t_0) = \frac{1}{2} + \frac{1}{\pi} \arctan\left(\pi \frac{t_0 - \bar{t}}{\delta t_0}\right) \quad (4)$$

where $N(x, \mu, \sigma)$ gives the normal distribution with mean $\mu$ and variance $\sigma^2$, and $L(x, \bar{x}, \delta x)$ gives a distribution with an approximate lower limit of $\bar{x}$ with some "slop" $\delta x$ (this distribution is not normalized). We consider the HST measurements, which give $\bar{H} \pm \delta H_0 = 80 \pm 17$ km/sec/Mpc, and a lower limit on the age of the universe of $\bar{t} = 11.5$ Gyr (Chaboyer 1994) with $\delta t_0 \simeq 0.5$ Gyr.

The age and Hubble constant are related to the cosmological parameters in an matter-dominated FRW universe by

$$\begin{aligned} H_0 t_0 &= 1.02 h \frac{t_0}{10 \text{ Gyr}} = f(\Omega, \Lambda) \\ &= \int_0^1 dx \, \frac{1}{\sqrt{1 - \Omega - \Lambda + \Omega x^{-1} + \Lambda x^2}} \end{aligned} \quad (5)$$

where $H_0 = 100h$ km/s/Mpc, $\Omega$ is the present density of non-relativistic matter and $\Lambda$ is the present vacuum density in units of the critical density (see, for example, Kolb & Turner 1989). This simplifies to the usual familiar forms in the case of $\Lambda = 0$. In particular, $H_0 t_0 = 2/3$ for $\Omega = 1, \Lambda = 0$.

In this case, we shall examine the following "theories," or classes of models:

1. $\Omega = 1, \Lambda = 0$;

2. $\Omega + \Lambda = 1$;

3. $0 \leq \Omega \leq 1, \Lambda = 0$;

4. $0 \leq \Omega \leq 1, 0 \leq \Lambda \leq 1$.

Model 1 has no parameters (and so is the "simplest"), models 2 and 3 both have 1 parameter, and model 4 has two parameters. If we chose one of the models with a parameter, we may of course use the usual Bayesian techniques to find confidence intervals for $\Omega$ and $\Lambda$. For now, we are not allowing for the possibility that $\Omega \geq 1$.) We have also not allowed any "cosmic variance," that is, the possibility that we live in a large underdense region and we are only measuring the local expansion rate.

Now, we must combine this information using Bayes' theorem and the usual rules from the calculus of probabilities. The parameters, before marginalization, are all of $H_0, t_0, \Omega, \Lambda$, so

$$\begin{aligned} p(H_0 t_0 \Omega \Lambda | DI) &= \frac{p(H_0 t_0 \Omega \Lambda | I)}{p(D|I)} p(D|H_0 t_0 \Omega \Lambda) \\ &= \frac{p(H_0 | t_0 \Omega \Lambda I) p(t_0 | \Omega \Lambda I) p(\Omega \Lambda | I)}{p(D|I)} p(D|H_0 I) \\ &= \frac{\delta\left(H_0 - f(\Omega, \Lambda)/t_0\right) L(t_0) p(\Omega \Lambda | I)}{p(D|I)} N(H_0, \bar{H}, \delta H_0), \end{aligned} \quad (6)$$

Now, we marginalize over $H_0$ and $t_0$:

$$p(\Omega \Lambda | I) = \frac{p(\Omega \Lambda | I)}{p(D|I)} \int dt_0 \, L(t_0) N\left[f(\Omega, \Lambda)/t_0, \bar{H}, \delta H_0\right] \quad (7)$$

Note that the integral in this expression is $p(D|fI)$, the likelihood for $f = H_0 t_0$ itself, or for $\Omega$ and $\Lambda$ through $f(\Omega, \Lambda)$. The result of this integral is a distribution cutting off to zero for $f \lesssim 80$ km/sec/Mpc $\times 11.5$ Gyr = 0.92, and rising roughly proportional to $f$ for $f \gtrsim 0.92$. Because of normalization problems, this distribution is difficult to calculate exactly; we will approximate it as

$$p(D|fI) \simeq fL\left(f, \bar{H}\bar{t}, \bar{H}\delta t_0 + \bar{t}\delta H_0\right). \quad (8)$$

This expression is correct in the limit $\delta H_0 \to 0$, turning the normal distribution into a Dirac delta function. Naively, we might expect the likelihood for $f$ to be this function without the factor of $f$ in front; since the observations favor large $f \approx 1$, the distributions are numerically very similar.

Finally, then, we have the posterior distribution

$$p(\Omega \Lambda | DI) = \frac{p(\Omega \Lambda | I)}{p(D|I)} f(\Omega, \Lambda) L\left(f(\Omega, \Lambda), \bar{H}\bar{t}, \bar{H}\delta t_0 + \bar{t}\delta H_0\right). \quad (9)$$

The different models we have enumerated above correspond to different values of the prior

$$p(\Omega\Lambda|I) = \begin{cases} \delta(\Omega - 1)\delta(\Lambda) & \text{model } 1 \\ \delta(\Omega + \Lambda - 1) & \text{"} \quad 2 \\ \delta(\Lambda) & \text{"} \quad 3 \\ 1 & \text{"} \quad 4 \end{cases} \quad (10)$$

For a given model, we can now plot the posterior, or we can integrate over the posteriors to compare the models. In Figures 1-3, we show the unnormalized posterior for cases 2-4. The integral over the posterior needed for setting the odds is given in Table 1. The "best fit" model is $\Omega + \Lambda = 1$; the worst is the simplest with $\Omega = 1$. The "most complicated" model 4 does quite well, but Figs. 3 shows that this is largely due to the large likelihood in the region with $\Omega + \Lambda \gtrsim 1$; if we instead use a prior $p(\Omega\Lambda|I) = 2\Theta(\Omega + \Lambda - 1)$, defining model $4'$, the odds drop to 1:2 with respect to model 1.

| Model | | Odds |
|---|---|---|
| 1 | $\Omega = 1, \Lambda = 0$ | 1 |
| 2 | $\Omega + \Lambda = 1$ | 7.51 |
| 3 | $\Omega \leq 1, \Lambda = 0$ | 2.66 |
| 4 | $\Omega \leq 1, \Lambda \leq 1$ | 5.96 |
| $4'$ | $\Omega + \Lambda \leq 1,$ | 2.09 |

Table 1: Odds ratios for various cosmological models

In general, the data favor a low $\Omega$ and/or a large $\Lambda$ as expected from the data which prefer $f \gtrsim 0.92$. For the two one-parameter models, we can calculate limits on these parameters, which are shown in Table 2 for confidence levels of 68%, 95% and 99%, the usual 1-, 2-, and 3-sigma levels. Obviously, parameters very close to the $\Omega = 1, \Lambda = 0$ point are ruled out, but not much more strongly than the prior information had already disfavored them.

| Significance | $\Omega < 1, \Lambda = 0$ | $\Omega + \Lambda = 1$ |
|---|---|---|
| 68% | $\Omega < 0.38$ | $\Lambda > 0.74$ |
| 95% | 0.87 | 0.30 |
| 99% | 0.97 | 0.07 |

Table 2: Confidence Intervals for $\Omega$ and $\Lambda$.

So far, we have used a fairly conservative prior range for $t_0$ in these calculations, only requiring that $t_0 \gtrsim 11.5$ Gyr. A less conservative (with respect to cosmology, not stellar evolution) limit might be $t_0 \gtrsim 14$ Gyr, perhaps with a larger "slop" $\delta t_0 \simeq 2$ Gyr. When we perform the calculation with these values, *none* of the models easily produce values of $f(\Omega, \Lambda) > 1$, so the posterior odds ratios relative to $\Omega = 1$, $\Lambda = 0$ slightly decrease. If we make other possible changes, such as using the CFHT results of $H_0 = 87 \pm 7$ km/s/Mpc, the ordering of the theories in relative likelihood remains the same: $\Omega + \Lambda = 1$ remains the most likely theory and $\Omega = 1$ the least. Since we have only approximated the likelihood function for $f$ in Eq. (8), it is good to see that the results are not too strongly dependent on the details of the procedure.

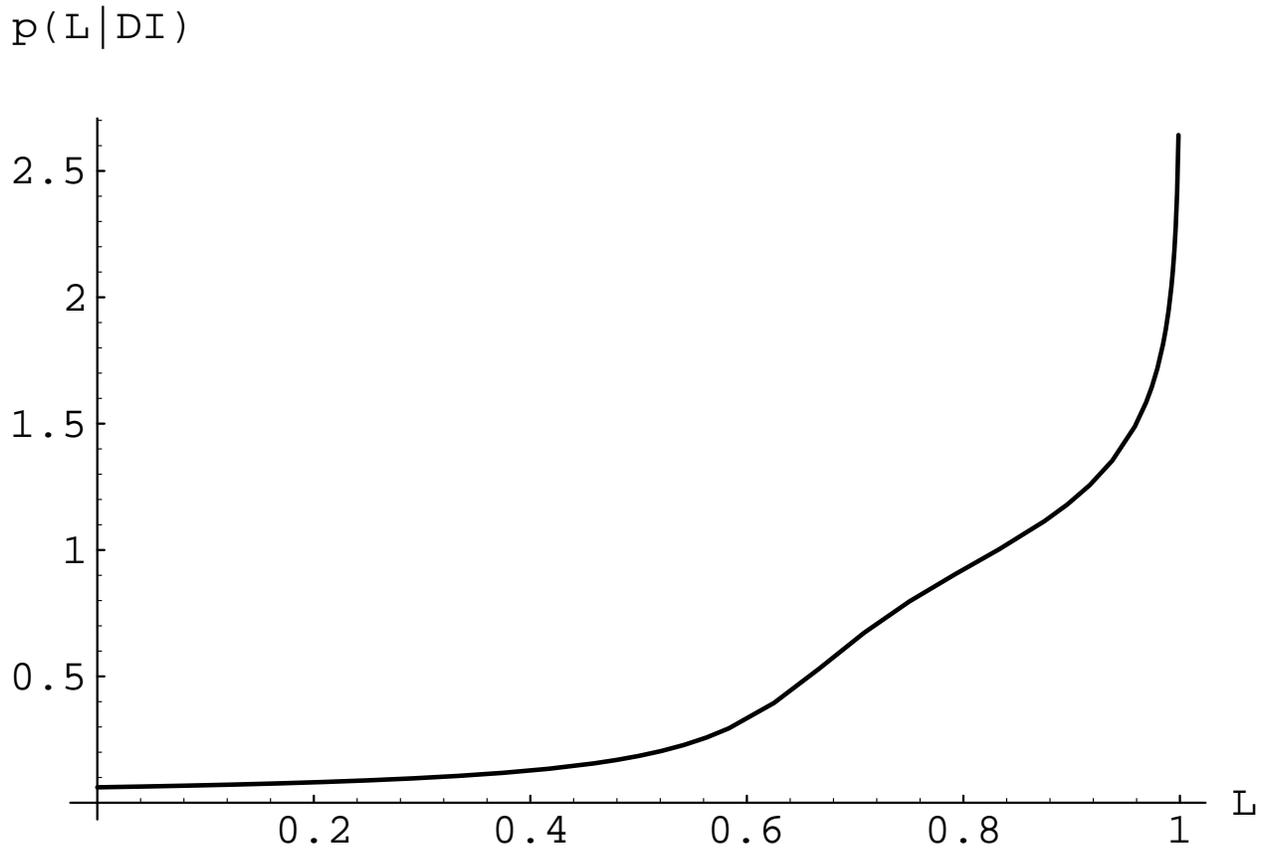

Fig. 1.— Unnormalized posterior distribution for $\Lambda$ for Model 2: $\Omega + \Lambda \leq 1$.

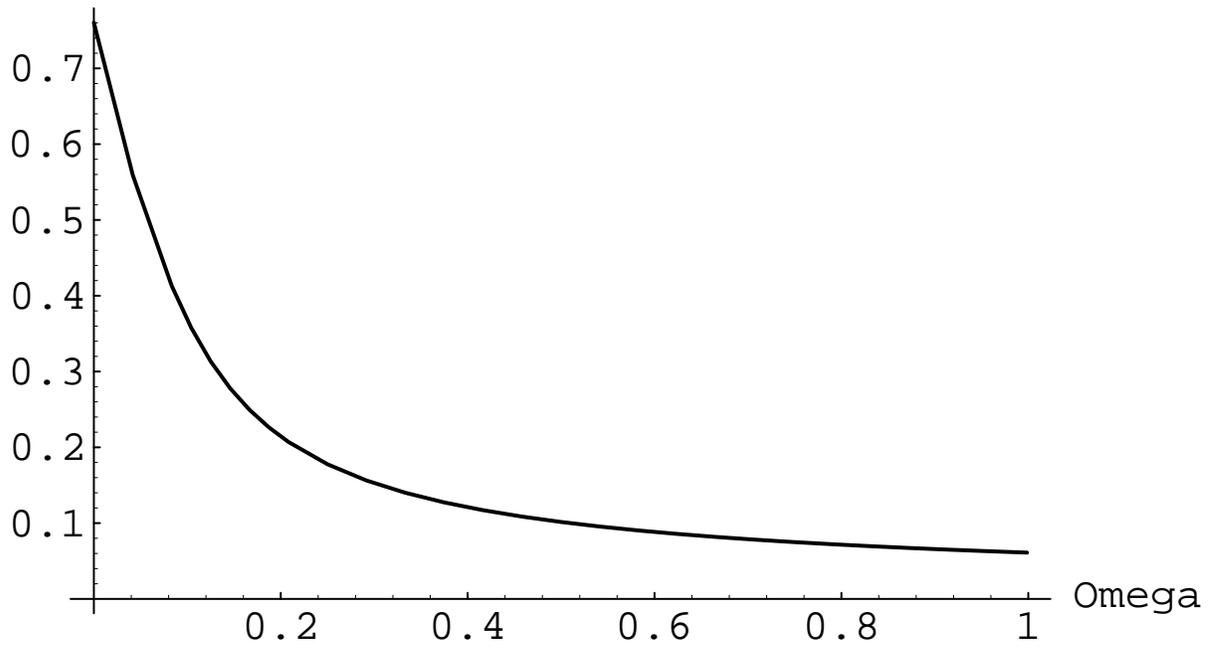

Fig. 2.— Unnormalized posterior distribution for $\Omega$ for Model 3: $\Omega \leq 1$, $\Lambda = 0$.

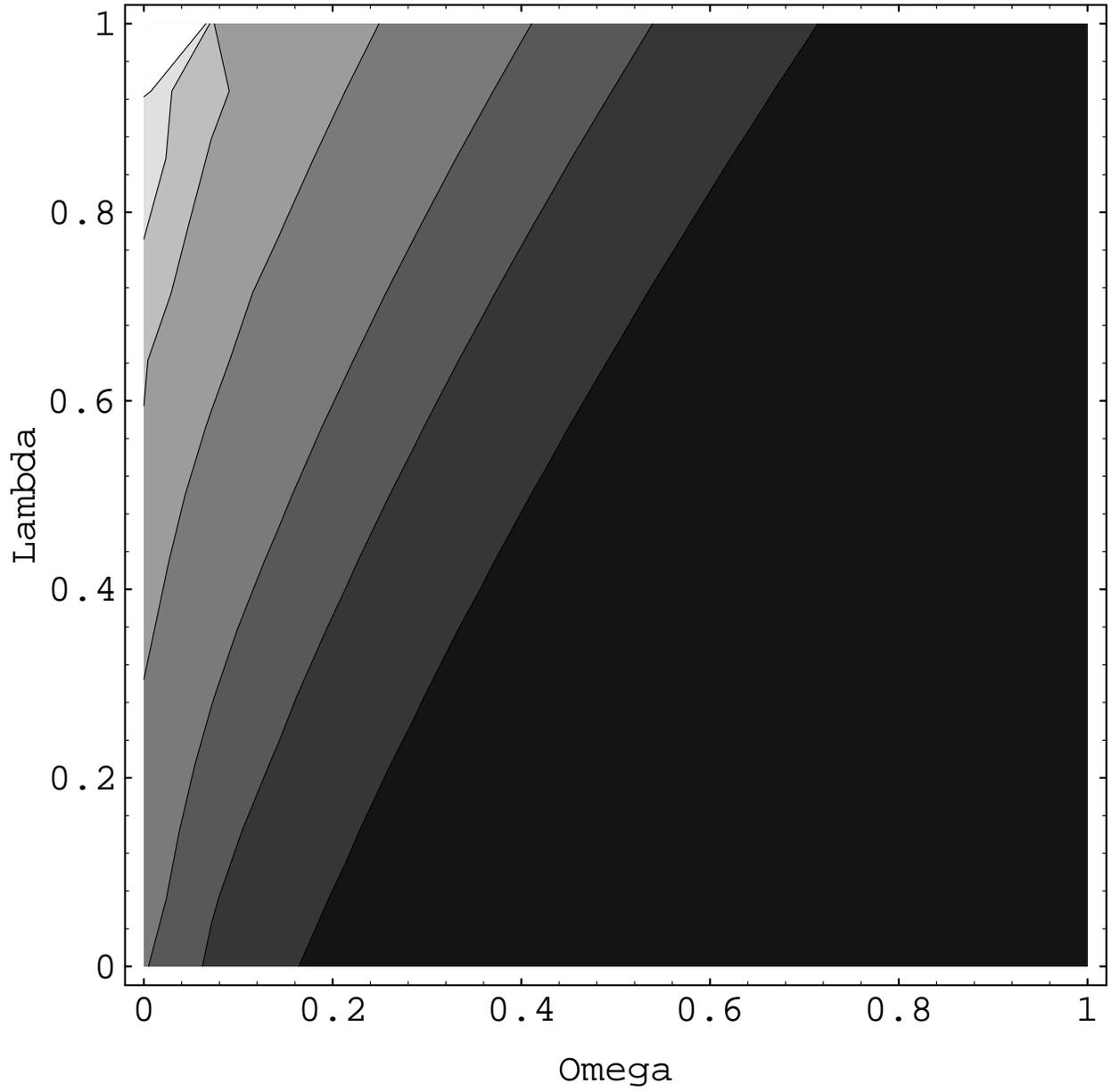

Fig. 3.— Unnormalized posterior distribution for $\Omega$ and $\Lambda$ for Model 4: $\Omega, \Lambda \leq 1$; contour spacing is 0.25 and lighter shades are more likely.

So what, in the end, does all this mean? We have quantified the conventional wisdom: the "best fit model" is one with $\Omega + \Lambda = 1$, favoring a cosmological constant $\Lambda \gtrsim 0.7$. Alternately, a low $\Omega$ is possible, although less strongly favored, requiring $\Omega \lesssim 0.4$, at least with current data. Recently, Leonard & Lake 1995 have performed a similar analysis, without the explicit emphasis on probabilistic methods, coming to the similar conclusion that $\Omega < 1$ and $\Lambda > 0$ is the most likely interpretation of the current data. As our knowledge of the age of the universe and the Hubble constant increase (if the central values we have used here are indeed correct), cosmological parameters further from the simplest flat universe with no cosmological constant will be required. One novel feature of this analysis is its "automatic" use of Ockham's razor; for example, the model with $\Omega + \Lambda \leq 1$ is disfavored with respect to the one-parameter models, even though it includes them as a subset, because *on average*, it predicts smaller values for $H_0 t_0$.

The author would like to thank Brian Chaboyer and Nick Kaiser for helpful discussion.